\documentclass[twocolumn,preprintnumbers,amsmath,amssymb,superscriptaddress]{revtex4-1}
\usepackage{graphicx}
\date {\today}

\begin{document}

\begin{abstract}

The model of the current paper is an extension of a previous publication, wherein we used the leaky integrate-and-fire model on a regular lattice with periodic boundary conditions, and introduced the temporal complexity as a genuine signature of criticality. In that work, the power-law distribution of neural avalanches was manifestation of supercriticality rather than criticality. Here, however, we show that continuous solution of the model and replacing the stochastic noise with a Gaussian zero-mean noise leads to the coincidence of power-law display of temporal complexity and spatiotemporal patterns of neural avalanches at the critical point. We conclude that the source of inconsistency may in fact be a numerical artifact originated by the discrete description of the model, which may imply slow numerical convergence of avalanche distribution compared to temporal complexity.

\end{abstract}
\author{Mohammad Dehghani Habibabadi}
\affiliation{Physics Department, Isfahan University of Technology, Isfahan, Iran (84156-83111).}
\author{Marzieh Zare}
\affiliation{School of Computer Science, Institute for Research in Fundamental Science, Tehran, Iran (19395-5746)}
\author{Farhad Shahbazi}
\affiliation{Physics Department, Isfahan University of Technology, Isfahan, Iran (84156-83111).}
\affiliation{School of Physics, Institute for Research in Fundamental Science, Tehran, Iran (19395-5531)}
\author{Javad Usefie-Mafahim}
\affiliation{Physics Department, Shahid Beheshti University, Tehran, Iran (19839)}
\author{Paolo Grigolini}
\affiliation{Center for Nonlinear Science, University of North Texas, Denton, TX, USA (76203)}
\title{Neuronal Avalanches: Where Temporal Complexity and Criticality Meet}
\maketitle

\section{introduction}
Diverse fields of research including neurophysiology, sociology, geophysics, and economics considers heavy tails and scale-free distributions as the signature of complexity \cite{Laherr}. The brain as a complex dynamic system and brain activities including neuronal membrane potentials, noninvasive electroencephalography (EEG), magnetoencephalography (MEG), and functional magnetic resonance imaging (fMRI) signals observed at many spatiotemporal patterns exhibit power-law behavior \cite{He2010}. 

Neuronal avalanches are a consequence of bursting activity of neurons in intervals less than a specific time. At critical state, avalanche size and lifetime distributions display power-law exponents limited to a specific range \cite{Beggs2003}. The widely varying profile of neural avalanche distribution in size and duration are described by a single universal scaling exponent, $\alpha$, in $ P(X) \sim X^{-\alpha}$. 

Criticality in neuronal avalanches is supported by existence of power-law exponent and shape collapse which are necessary criteria for criticality as proved in cultured slices of cortical tissue \cite{Friedman2012}. The existence of power-law distributions in neuronal avalanches was detected in cortex slice cultures of rats in vitro \cite{Shew2009,Yang2012}, rat cortical layer 2/3 at the beginning and end of the second week postnatal \cite{Gireesh2008}, Local Field Potentials (LFPs) of anesthetized cats \cite{Hann2010, Priesemann2014}, new born rats \cite{Yang2012}, and monkeys \cite{Petermann2009,Priesemann2014}.
According to the popular view of Per Bak et. al., \cite{Bak}, criticality is realized spontaneously by complex systems rather than requiring the fine tuning of a control parameter with a critical value. They proposed the term Self-Organized Criticality (SOC), as an operation of the system at criticality that generates the power law behavior in natural phenomena. 
There are widely empirical evidences showing that the brain works near criticality \cite{ Beggs2012, Friedman2012,Yang2012,Socolar2003, Shew2009}. Furthermore, many studies have shown that SOC in the brain is often accompanied by other optimal operations including phase synchrony \cite{Yang2012}, information storage \cite{Socolar2003}, communication and information transition \cite{Beggs2003}, Optimal Communication \cite{Beggs2003, Bertschinger2004, Ramo2007, Beggs2012, Shew2011, Tanaka2009, Chialvo2010}, transition capability \cite{Shew2011}, computational power \cite{Bertschinger2004}, and dynamic range \cite{Shew2011,Kinouchi2006}.

Power law distribution may suggest that neural networks operate near a non-equilibrium critical point where the phase transition occurs \cite{Zare2013}. A critical point sets a boundary between an ordered and a less ordered state with different scaling behaviors \cite{Friedman2012, Zare2013}. However, since many possible mechanisms generate power law, the appearance of them alone is insufficient to establish criticality, and also non-critical systems may produce power laws \cite{Beggs2012,PLfake1,PLfake2,Newman2005}. In \cite{Destexhe2010, Dehghani2010}, authors argue that the methodology used in the analysis can amplify or create power-laws that are not related to critical structure in the underlying signal.

However, in a numerical study~\cite{Zare2013}, using a form of integrate-and fire model, we have shown that \textit{temporal complexity} is a robust indicator of criticality. We found that temporal complexity occurs in a narrow range of values of the control parameter. Using cross-correlation and mutual information measures, we showed that information transfer from one network to another becomes maximal in the corresponding range -- Note that the two networks were identically selected as regular lattice, similar to the network of the current study. We argued that if the enhancement of information transfer is interpreted as a signature of criticality, then the power law avalanches are a manifestation of supercriticality rather than criticality. This finding, on the other hand, read well with the finding of \cite{Dehghani2012} that no clear evidence of power-law scaling or self-organized critical states was found in the awake and sleeping brain of mammals including monkey, cat and human.

However, this result, would not conflict with the hypothesis that the brain works near criticality, as far as temporal complexity--decoupled from the density distribution of avalanches -- is an indicator of criticality. Yet, considering many inconsistent results reported from recordings in awake animals and humans in appearance of power-law scaling of avalanche data profile, we assumed that it may be a conflict that avalanche distributions in size and time do not display power-law behavior at the critical point indicated by temporal complexity, and we did not cease our attempt to find the source of the conflict. 

Hence, in this study, we show that by replacing the discrete stochastic noise with a continuous Gaussian noise, avalanche probability distributions display power law behavior at criticality indicated by temporal complexity, and assume that the choice of the noise may be the source of the conflict between our previous study \cite{Zare2013}, and the current study. 

Therefore, we first show that temporal complexity detects a critical point at the phase transition, where it is poised between a random and a regular state. We then examine whether avalanche distribution in size and time follow a power-law exponent as found in \cite{Beggs2003}. Using scaling theory \cite{James} that predicts exponent relations, we confirm that neural avalanche data collapses into critical exponent suggested by temporal complexity approach. 

\section{ The Survival Probability as the Mittag-Leffler Function}

In \cite{ml}, Metzler and Klafter have emphasized the importance of the Mittag-Leffler (ML) survival probability. ML function settles a bridge between the stretched exponential and inverse power law survival probabilities as important signs of complexity. ML function is the generalized exponential function \cite{ml}:

\begin{equation} \label{MLmain}
\ E_{\alpha}(Z) = \sum_{n=0}^{\infty} \frac{Z^n} {\Gamma(1+ \eta \alpha)},
\end{equation}
with $\alpha$ being an arbitrary positive real number. In the case of $\alpha< 1$, this function is interpreted as a survival probability in time and it is frequently written as $E_{\alpha}(-(\lambda t)^\alpha)$. In the short time region, $t < 1/\lambda $ the ML survival probability is described by the stretched exponential function. 
\begin{equation} \label{property1}
\ E_{\alpha}(-(\lambda t)^{\alpha}) \propto \exp(-(\lambda t)^{\alpha}).
\end{equation}
and in the large time region $t > 1/\lambda $
by
\begin{equation} \label{property2}
\ E_{\alpha}(-(\lambda t)^{\alpha})\propto \frac{1}{t^{\alpha}}.
\end{equation}

The concept of survival probability is connected to the theoretical perspective of a complex system
generating events in time. A complex system at criticality generates events that are referred to as critical events. According to a coin tossing prescription \cite{Zomfon}, the time interval between two consecutive crucial events, i.e. laminar region, is assigned the values $\pm 1$. At time $t = 0$, the system is prepared by selecting all the realizations with an event occurring at that time and positive laminar regions. The probability that no event occurs up to time $t$ is properly termed survival probability, denoted by $\Psi(t)$, and
the function $\psi(t) \equiv d\Psi (t)/dt $ is called waiting time density. We adopt the symbols $\Psi_{ML}(t)$ and $\psi_{ML}(t)$ to denote the ML survival probability and the corresponding waiting time density function, respectively.

In a study on EEG dynamics of the human brain \cite{Bianco}, the stretched exponential of Eq.\ref {property1} was interpreted as the top of a ML iceberg, with the inverse power law hidden below the sea surface.
A strong support to the conjecture of a connection between the ML function and criticality is given by the discovery that the ML function is universal \cite{Pensri}. In the literature, complexity is supposed to be associated to the deviation from the exponential prescription through an inverse power law structure. For instance, a plausible form for the time distance between two criticality-induced consecutive events may correspond to the survival probability 
\begin{equation} \label{visible}
\Psi(t) = \left(\frac{T}{T + t}\right)^{\alpha}.
\end{equation}
with $\alpha< 1$. 
The process is thought to be complex because for times $t \gg T$, it is identical to the non integrable inverse
power law $1/t^{\alpha}$. No attention is paid to the time region $t \approx T$, which does not even show up if a log–log representation is adopted. However, the temporal complexity of systems generated by the cooperation of a finite number of components \cite{Gosia} is characterized by an exponential truncation at long times.

In this article, we demonstrate emergence of criticality indicated by survival probability distribution, termed temporal complexity analysis. The distribution does fit well with ML function. To this end, we set a time limit to the inverse power law regime, and also pay attention 
analysis of the short-time regime to establish even tenuous signs of cooperation which is the system's control parameter. We refer to 
Eq.\ref{visible} as an example of analytical representation of temporal complexity where
the analysis of the short-time behavior cannot disclose the temporal complexity of the process, clearly emerging
in the long-time limit through the inverse power law $1/t^\alpha$: no clear signs of the power index $\alpha$ may emerge from the short-time analysis. The same argument applies to the ML complexity if the stretched-exponential regime of Eq.\ref{property1} is not extended enough in time. 

We assign $\hat{\Psi}_{ML} (u)$ to the Laplace transform of ML survival probability, by adopting the notation for the Laplace transform of $(\hat{\Psi}(u)= \int_0^\infty dt \Psi(t) e^{-ut})$, on Eq.\ref{MLmain}

\begin{equation} \label{laplaceml}
\ \hat{\Psi}_{ML}(u)= \frac {1}{u+\lambda^{\alpha} (u+\Gamma_t)^{1-\alpha}}
\end{equation}

\noindent for $ \alpha<1$, $\Gamma_t=0$. When $1/\lambda$ is of the order of the time step and $1/\Gamma_t $ is much larger than the unit time step, the survival probability turns out to be virtually an inverse power law, whereas when $1/ \lambda$ is of the order of $1/\Gamma_t $ and both are much larger than the unit time step, the survival probability turns out to be a stretched exponential function. 
Failli et al. \cite{Failla} illustrate the effect of establishing a cooperative interaction in the case of the random growth of surfaces. A growing surface is a set of growing columns whose height increases linearly in time with fluctuations that, in the absence of cooperation, would be of Poisson type. The effect of cooperative interaction is to turn the Poisson fluctuations into complex fluctuations: the interval between two consecutive crossings of the mean value being described by an inverse power law waiting time distribution $\psi_{ML}(t) $, corresponding to a survival probability, whose Laplace transform is given by Eq. \ref{laplaceml}.

In conclusion, according to the earlier work \cite{Failla}, we interpret $\alpha<1$ as a manifestation of the cooperative nature of the process. In this research, we illustrate a neural model where the time interval between two consecutive firings, in the absence of cooperation is described by an ordinary exponential function, thereby corresponding to $\alpha=1$. The effect of cooperation is to make $\alpha$ decrease in a monotonic way, when increasing the cooperation strength, $K$. As done in the earlier work of \cite{Zare2013}, we establish the parameters $\alpha$, $\lambda$, $\lambda ^{\alpha}$ by relying on fitting the Laplace transform
of the numerical survival probability, i.e. function $\hat {\Psi}_{ML}(u)$.

\section{Model Description}
The model used in the previous works \cite{Zare2013, Zare2012,Zare2015}, and here is the leaky integrate-and fire model (LIFM) \cite{Politi}.
\begin{equation} \label{LIFM}
\dot{x_i} = - \gamma x(t_i) + S + \sigma \xi_i (t),
\end{equation}
where $x$ is the membrane potential, $1/\gamma$ is the membrane time constant of the neuron, $S$ is proportional to a constant input current. Note that $i = 1,..., N$ where $N$ is the total number of neurons. Each neuron starts from a random value or zero, and fires when it reaches the threshold, $x = 1$. When a neuron fires, it forces all the neurons linked to it to make a step ahead by the quantity $K$ which means that all neurons are excitatory. The parameter $K$ plays the all-important role of control parameter, and is expected to generate criticality when the special value $K_c$ is adopted. After firing, each neuron jumps back to the rest state $x = 0$. 

When $K = 0$, the vanishing noise condition yields the following expression for the time distance between two consecutive firings of the same neuron 
\begin{equation} \label{perfectperiodicity}
T_P = \frac{1}{\gamma} \ln \left(\frac{1}{1- \frac{\gamma}{S}}\right).
\end{equation}
If $K >0$, after a few time steps all the neurons fire at the same time \cite{Mirollo}, thereby generating a sequence of quakes of intensity $N$ with the time period $T_P$ given by Eq.\ref{perfectperiodicity}.

\noindent The parameter $\sigma$ is the standard deviation of the noise which can also be considered as the noise intensity. In the previous solution of Eq. \ref{LIFM}, we adopted the integration time step 
$\Delta t= 1$ and $\xi(t)$ as a discontinuous random fluctuation taking with equal probability either the value of $\xi(t)=1$ or $\xi(t)=-1$~\cite{Zare2013, Zare2012}. However, the central limit theorem requires that a discrete noise with finite variance converges to a Gaussian distribution after long time integrate. Hence, in order to have results comparable with real neuronal behavior, it is natural to have a continuous time description of the LIFM. Therefore, we treat the time continuously and then consider $\xi$ to be a continuous Gaussian white noise with zero mean and unit variance, defined by 

\begin{eqnarray} \label{noise}
\langle \xi(t) \rangle &=&0 \nonumber\\
\langle \xi(t) \xi(t') \rangle &=& \delta(t-t').
\end{eqnarray}

To reduce the number of the parameters, we define the dimensionless time variable ${\cal T}=\gamma t$, hence, equation Eq.\ref{LIFM} can be rewritten in the following form

\begin{equation}
\label{LIFM2}
\dot{x_i}({\cal T}) = -x_{i}({\cal T}) + \frac{S}{\gamma} +\frac{\sigma}{\sqrt\gamma} \eta_i ({\cal T}),
\end{equation}
in which $\eta({\cal T})=\frac {1} {\sqrt{\gamma}} \xi (\frac {\cal T} {\gamma})$ is the dimensionless Gaussian noise with zero mean and unit variance. To numerically integrate the stochastic differential equation Eq.\ref{LIFM}, we use the Ito's interpretaion, which is

\begin{equation}
x_{i}({\cal T}+d{\cal T})=x_{i}({\cal T})+[-x_{i}({\cal T}) + \frac{S}{\gamma}]d{\cal T}+ \frac{\sigma}{\sqrt\gamma} \eta_i ({\cal T}) \sqrt{d{\cal T}}.
\end{equation}

Here, we choose $d{\cal T}=0.01$, $S = 0.001005$, $\gamma = 0.001$, and $\sigma=0.0001$. We assume that neurons are residing on the nodes of a two-dimensional square lattice with periodic boundary condition with size $N=L \times L$, where $L$ is the linear size of the lattice (Here, $L = 10, 15, 20$). The duration of all realizations was $10^7$ time steps. 

The adoption of periodic boundary conditions is to ensure the total equivalence of the cooperating units, so as to avoid the doubt that the onset of firing bursts may be triggered by units with a favorable topology. However, numerical calculations not reported here show that the adoption of periodic boundary condition is not crucial for the results of this paper. Also, due to computational expenses, the number of neurons in the current research is limited to $N=400$. In \cite{Zare2012}, we have studied networks up to size $N=2500$, and have extensively discussed the finite size effect. 
\begin{figure}
\centering
\includegraphics[width=9cm, height=7cm]{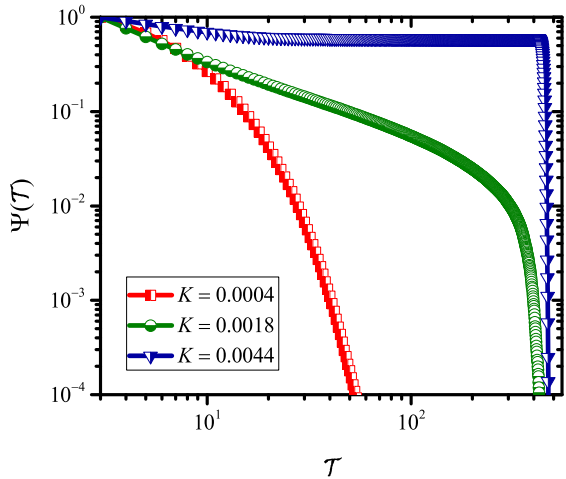}
\caption{(Color online) Survival probability $\Psi$ versus the dimensionless time ${\cal T}=\gamma t$, for three different cooperation parameters $K=0.0004, 0.0018, 0.0044$ with $\sigma = 0.0001$, $\gamma= 0.001$, $S = 0.001005$ in a $ 10\times 10$ lattice. }
\label{fig:f1}
\end{figure}
The time distance between two consecutive firings of a set of $N$ neurons, in the lack of cooperation is given by $G=\frac{N}{<\tau>}$ with $<\tau>\approx T_P$ and the dynamics is Poissonian process defined by an exponential
\begin{equation} \label{poisson}
\Psi(t) = e^{-Gt}.
\end{equation}
which indicates the probability that no firing occurs up to the time $t$ from an earlier firing. According to \cite{complexity}, cooperation generates scale invariance, hence, increasing $K$ leads to a transition from the exponential form of Eq.\ref{poisson} to $T$ so small as to make Eq. \ref{visible} virtually equivalent to the inverse power law of Eq.\ref{property2} over the available time scale. 

We noted that the exponential function of Eq. \ref{poisson} corresponds to the ML function with $ \alpha = 1$ and $\lambda = G$. This suggests that temporal complexity becomes evident when the parameter $\alpha$, as determined by means of the fitting procedure, becomes significantly smaller than 1.
\begin{figure}
\centering
\includegraphics[width=8cm, height=6.5cm]{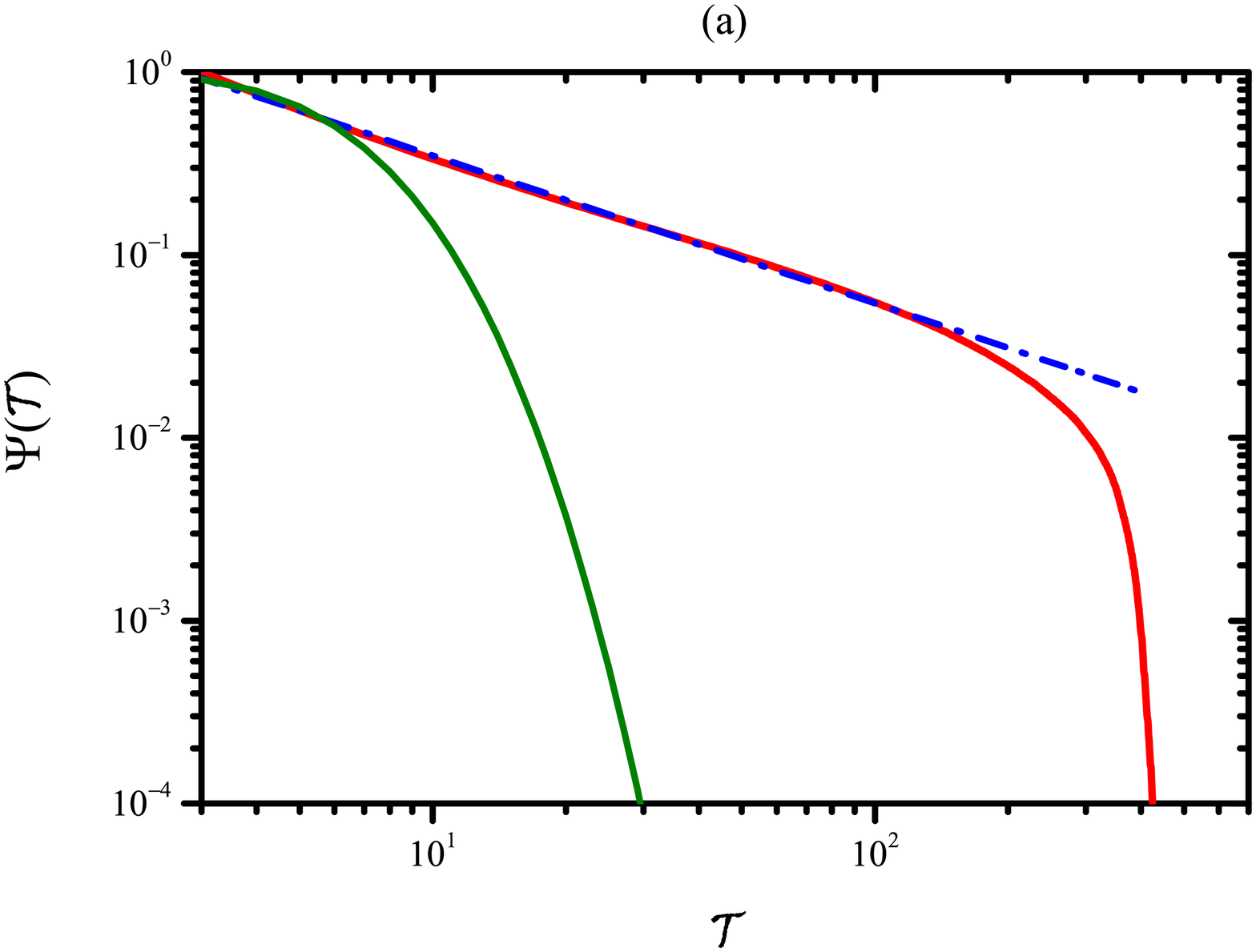}
\includegraphics[width=9cm, height=6.8cm]{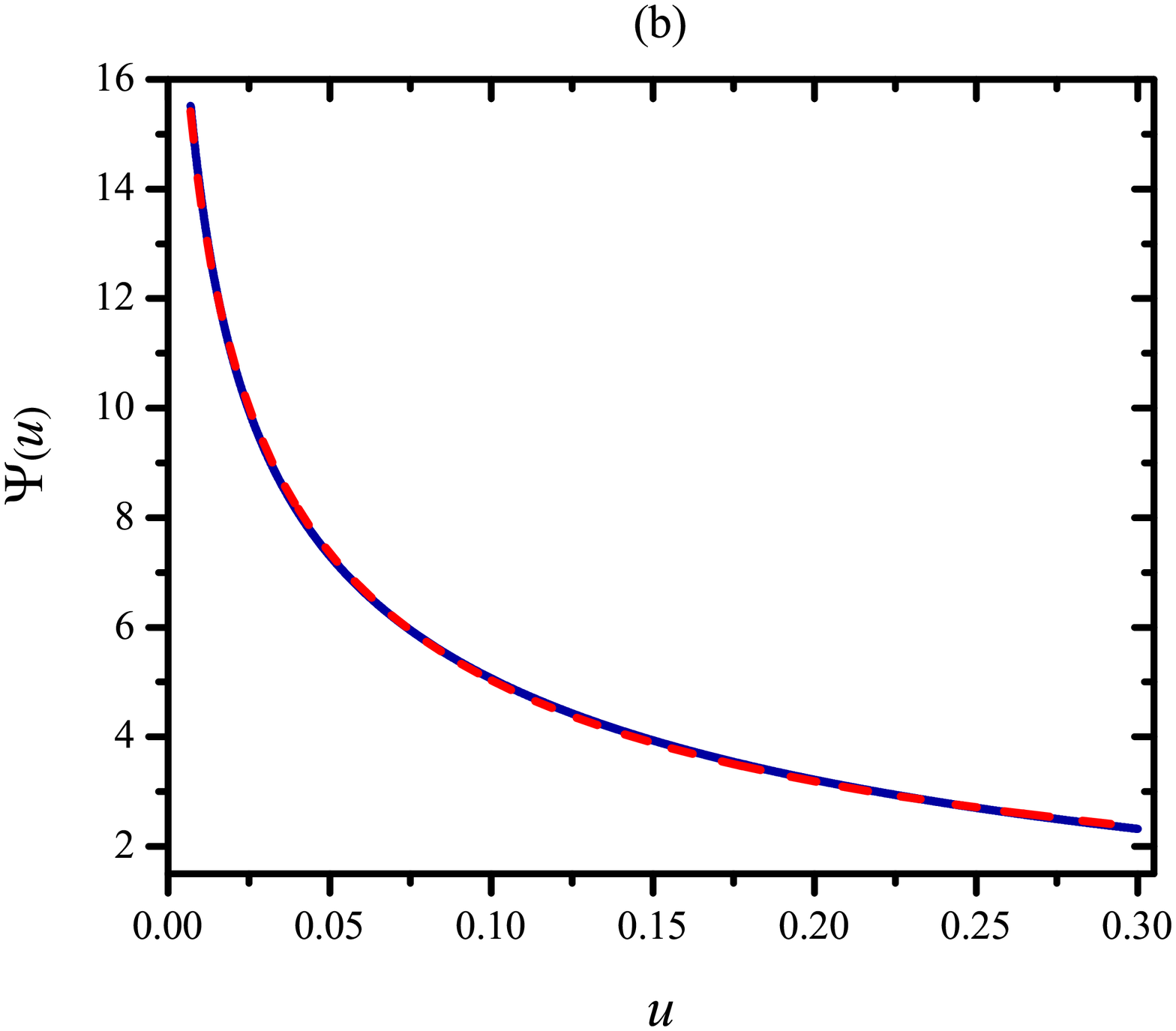}
\caption{(Color online) Survival probability $\Psi$ versus the dimensionless time ${\cal T}=\gamma t$, for $K=0.0018$ with $\sigma = 0.0001$, $\gamma= 0.001$, $S = 0.001005$ in a $ 10\times 10$ lattice.(a) Fitting of survival probability in time regime, The green curve is the stretched exponential $ \exp ({-\lambda t}^{\alpha})$ and the blue dash line is the inverse power law $1/t^\alpha$. (b) in Laplace regime. The results of numerical survival well fit into Eq.\ref{laplaceml}}
\label{fig:f2}
\end{figure}
Fig.\ref{fig:f1}. shows how neural network deviates from exponential to a regular behavior with changing control parameters, $K$. Two distinctive time regimes characterize the survival probability: short time and longtime regimes that can be fitted by stretched exponential $ \exp(-(\lambda t)^{\alpha})$ with $\alpha <1$ and inverse power law $1/t^{\alpha}$ respectively. Fitting parameters, $\alpha$ and $\lambda$ can be found under a fitting procedure. $\alpha$ indicates the power law exponent of survival probability holding $0<\alpha<1$, and $\lambda$ indicates the scale of the stretched exponential of survival probability distribution. 

As an example, fitting procedure for a value of $K=0.0018$, is shown in Fig.\ref {fig:f2}-(a). The procedure is accomplished on each time regime in two steps: fitting of the stretched exponential and fitting of the inverse power law. However, fitting on the Laplace transformation of survival probability can be done in a single step on the total observation time. It is important to note that the fitting parameters found in both cases are equal. 
\begin{figure}
\centering
\includegraphics[width=9cm, height=7cm]{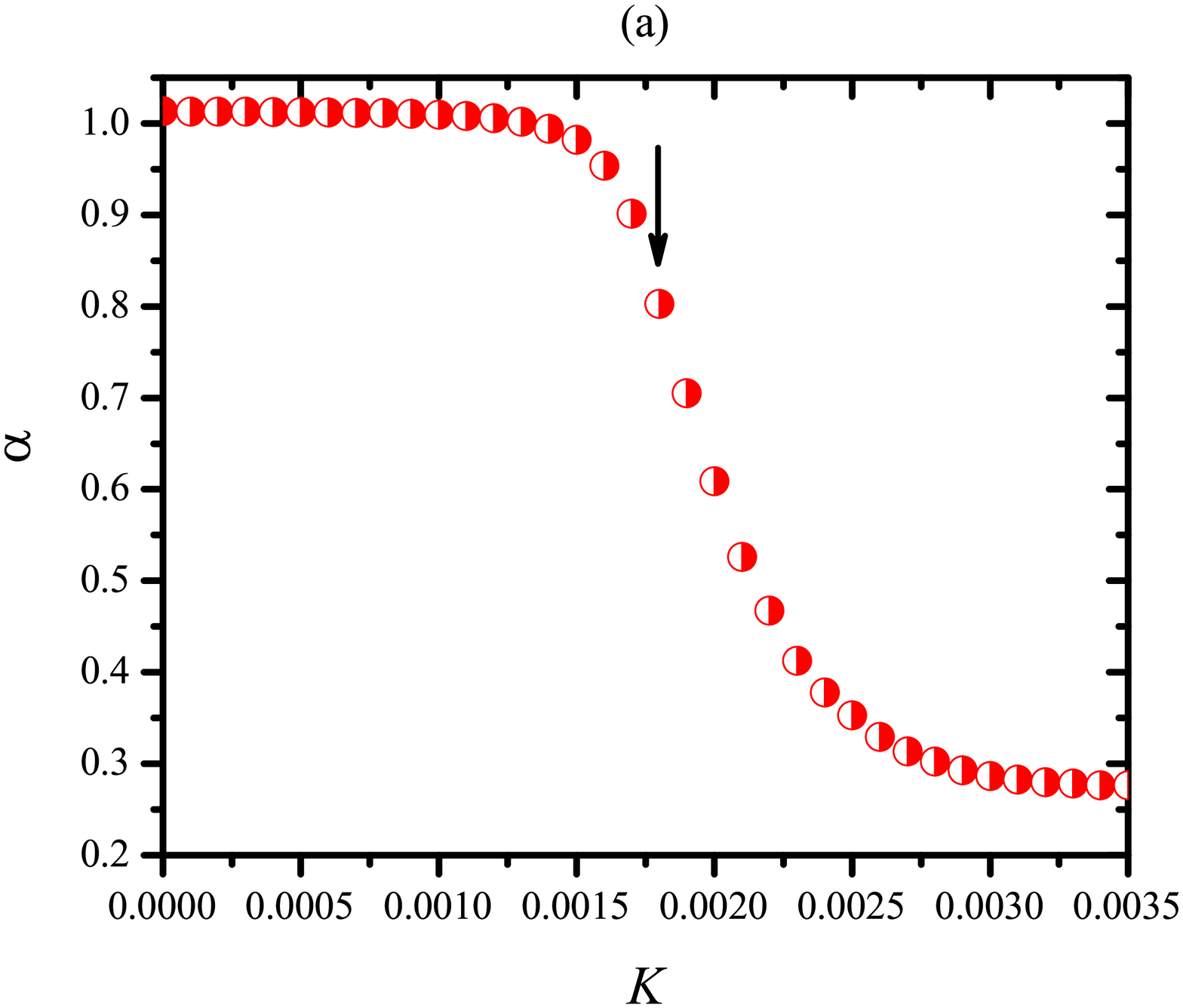}
\includegraphics[width=9cm, height=7cm]{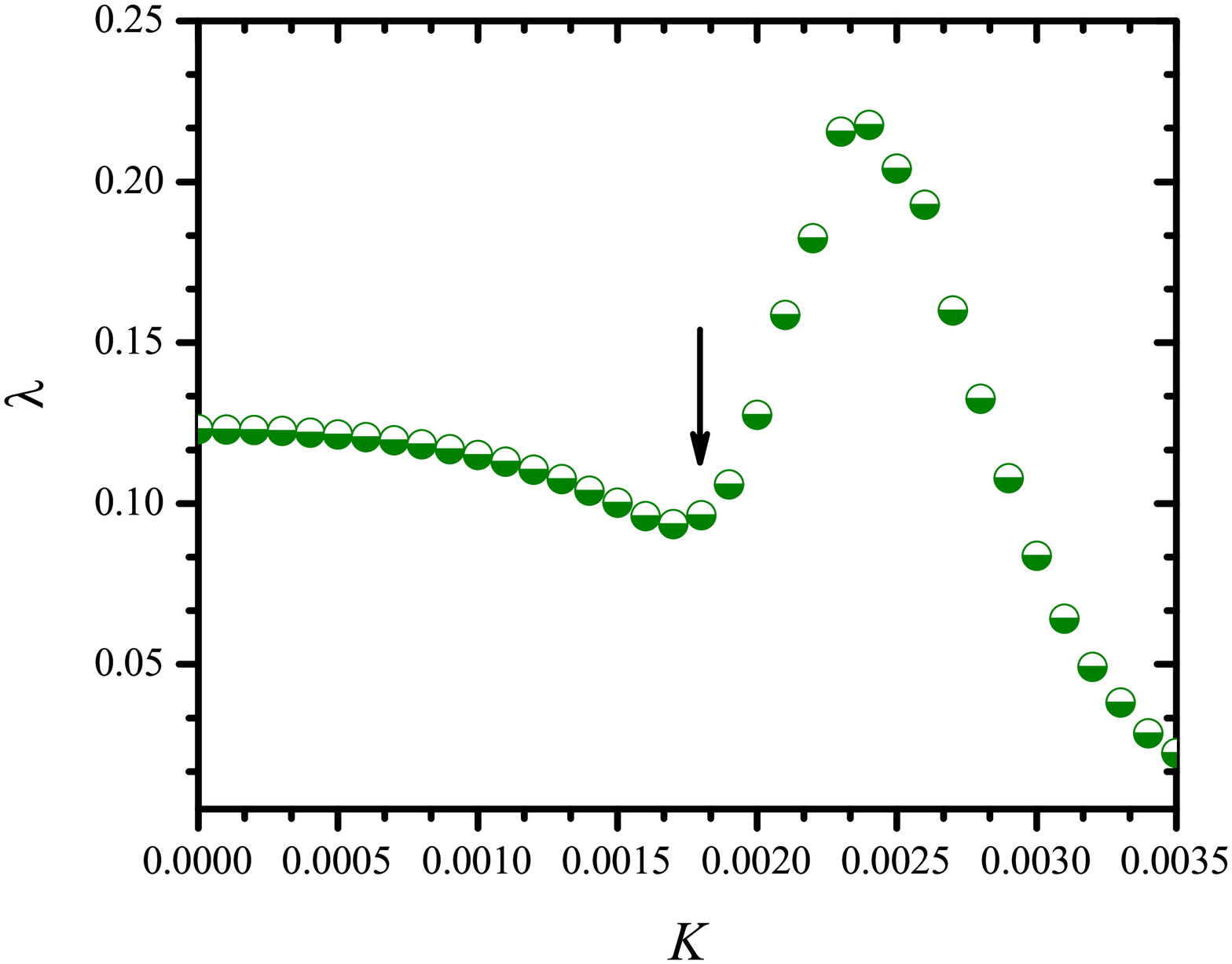}
\includegraphics[width=9cm, height=7cm]{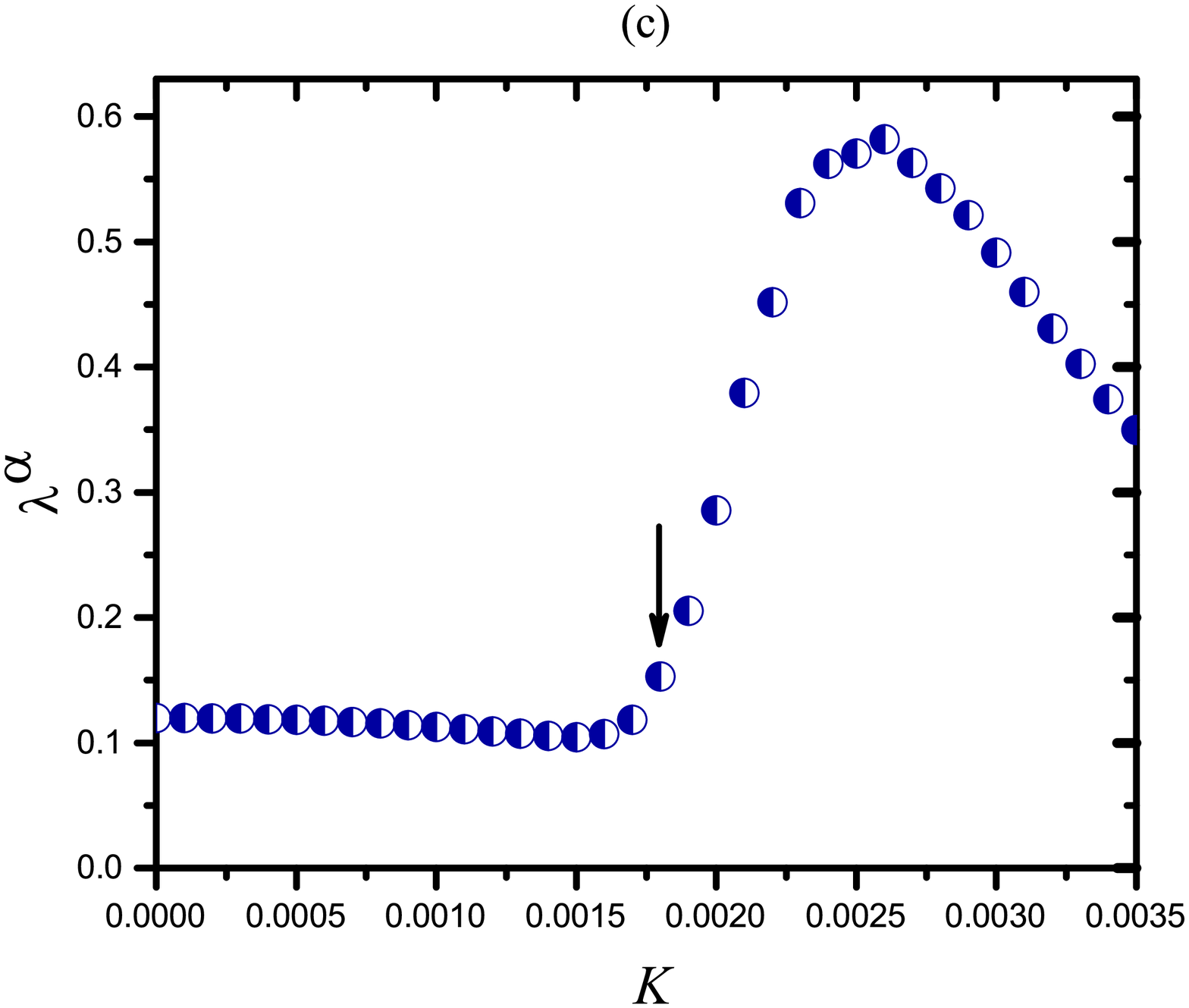}
\caption{(Color online) (a) Variation of the power index $\alpha$, (b) Scale of stretched exponential, $\lambda$ , and (c) $\lambda^\alpha$ defined by Eq.\ref{laplaceml}, versus the cooperation parameter, $K$. Notice the abrupt change in $K=0.0018$.}
\label{fig:f3}
\end{figure}

\begin{figure}[t]
\centering
\includegraphics[width=9cm, height=7cm]{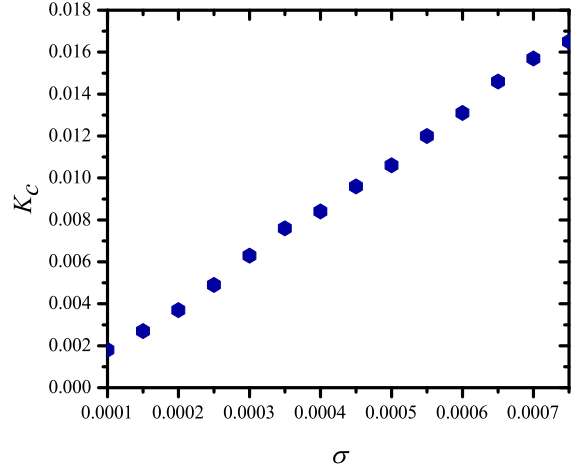}
\caption{ (Color online) Variation of the critical coupling parameter, $K_c$, 
versus noise intensity, $\sigma$, for $\gamma= 0.001$, $S = 0.001005$ and $N = 100$.}
\label{fig:f4}
\end{figure}

Hence, it is more convenient to use the Laplace transform of survival probability and find the fitting parameters. The procedure is as follows: i) apply a Laplace transform on survival probability gained from neural dynamics, ii) fit the data given in step (i) into Eq. \ref{laplaceml} to find $\alpha$ and $\lambda$. We repeat this procedure for different values of $K$. The Laplace transform of survival probability of the same $K$ is calculated as shown in Fig.\ref{fig:f2}-(b). 

The corresponding fractal exponents of each curve, $\alpha$, $\lambda$ and $\lambda^{\alpha}$ are plotted in Fig.\ref{fig:f3} respectively. We interpret the steepness of changes as criticality, which is visually inspected at $K=0.0018$. In the previous work \cite{Zare2013}, we confirmed the existence of criticality by aging experiment and information transfer, and as the continuous solution of the model did not conflict with the results of the previous work \cite{Zare2013}, we encourage readers to see that for more details.

The model of the current paper is a generalization of the model proposed by Mirollo and Strogatz \cite{Mirollo} in which in the absence of noise, full synchronization is achieved after few steps in the absence of noise. However, adding a noise to the model in order to generate temporal complexity along with an adjustable coupling parameter, $K$, a competition is set between these two parameters. We have explored how coupling has to be adjusted to the noise to maintain the criticality and phase transition. Hence, we run the model for different values of $\sigma$ and arrive at Fig.\ref{fig:f4}. As can be inferred from this figure, the critical coupling $K_c$, approximately has a linear dependence on the noise intensity, and as we increase the noise in the system, cooperation has to level up in order to maintain the system at criticality. 

Now, we focus on neural avalanches to explore whether this change results in appearance of power-law behavior of avalanche distribution at $K_c$ and ask if avalanche data including avalanche size, avalanche duration and temporal profile fall within SOC predictions according to \cite{Friedman2012, James}.

\section{Neural Avalanches}
In our previous study \cite{Zare2013}, we did not observe the coincidence of power-law distribution of neural avalanches at criticality which was realized by temporal complexity while confirmed by aging experiment and information transfer. This finding may cast doubt on the operation of the brain near criticality; however, we hypothesized that temporal complexity as a promising detector of criticality may suffice. 

Here, however, we explore whether the source of conflict is due to the discrete solution of the model, generating a numerical artifact. Hence, we calculate the size and duration of the neuronal avalanches creating as the results of our LIFM. For this purpose, we count the number of firing neurons in time bins of $5$ simulation steps ($\Delta {\cal T}=0.01$). An avalanche is recorded whenever a burst of neurons is followed by quiescent duration of minimum $5$ steps. The number of neurons fired during this active region is called the avalanche size, $S$, and the duration of this activity is called the avalanche duration $T$. 
\begin{figure}
\centering
\includegraphics[scale=0.35]{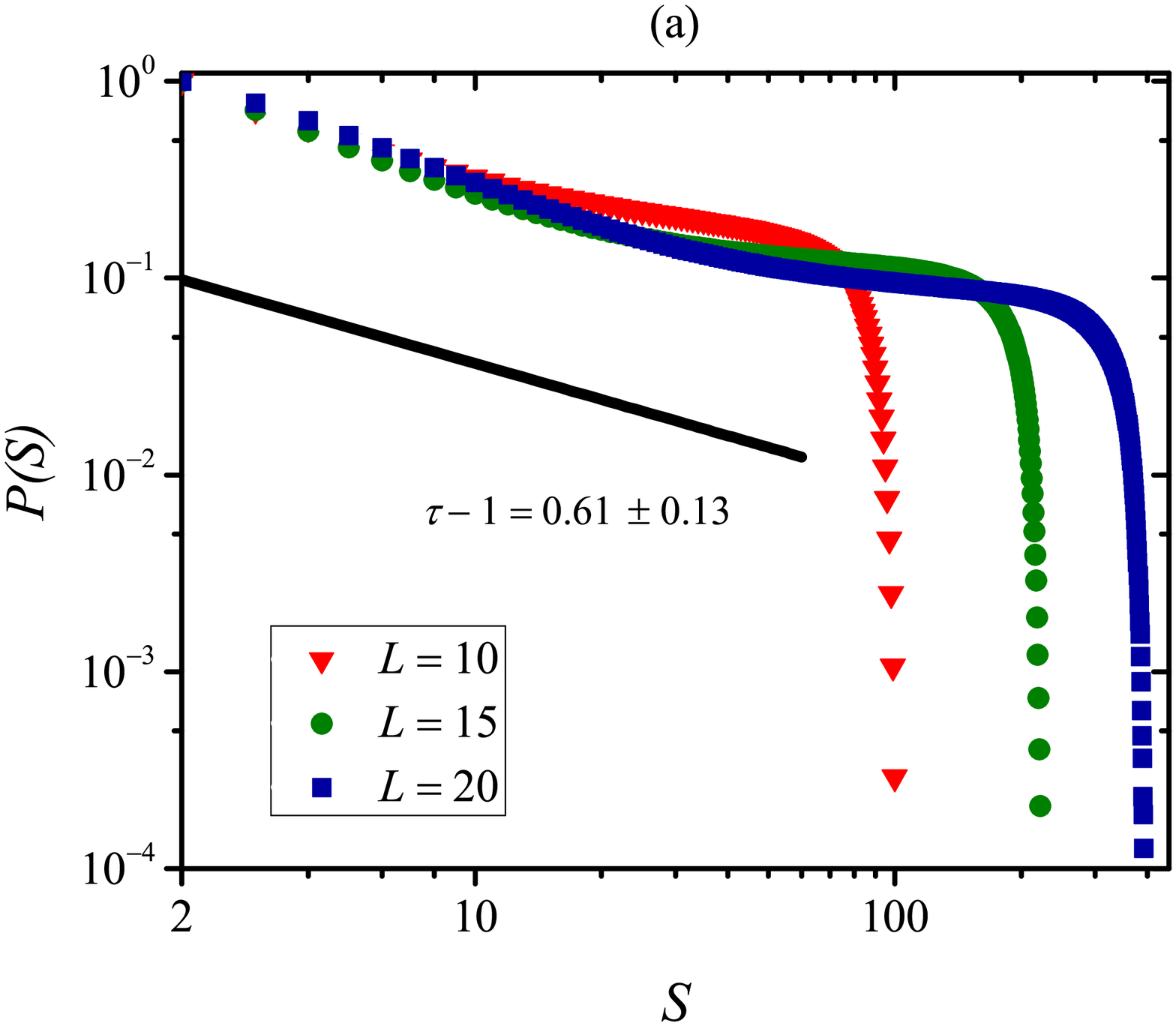}
\includegraphics[scale=0.35]{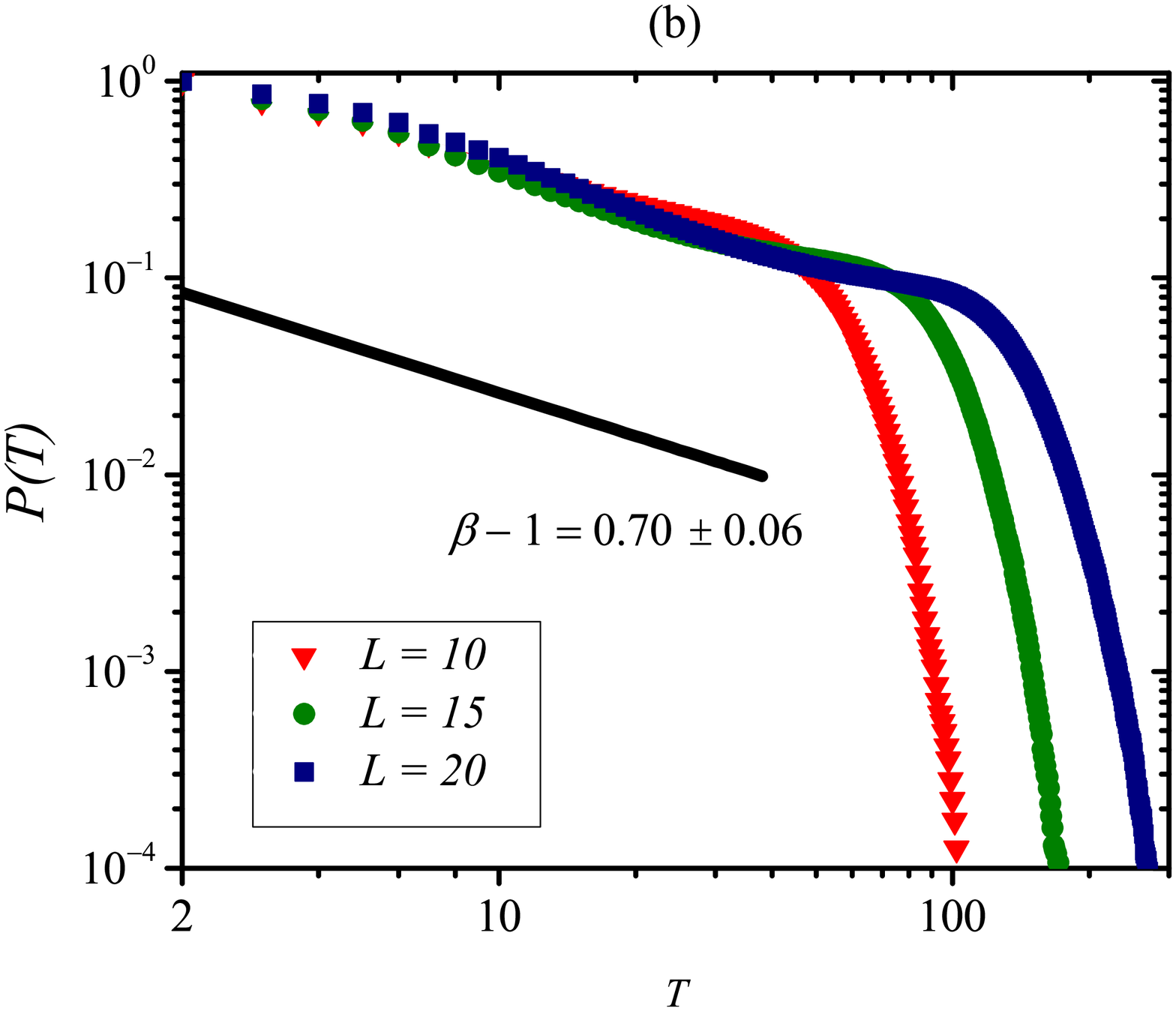}
\includegraphics[scale=0.32]{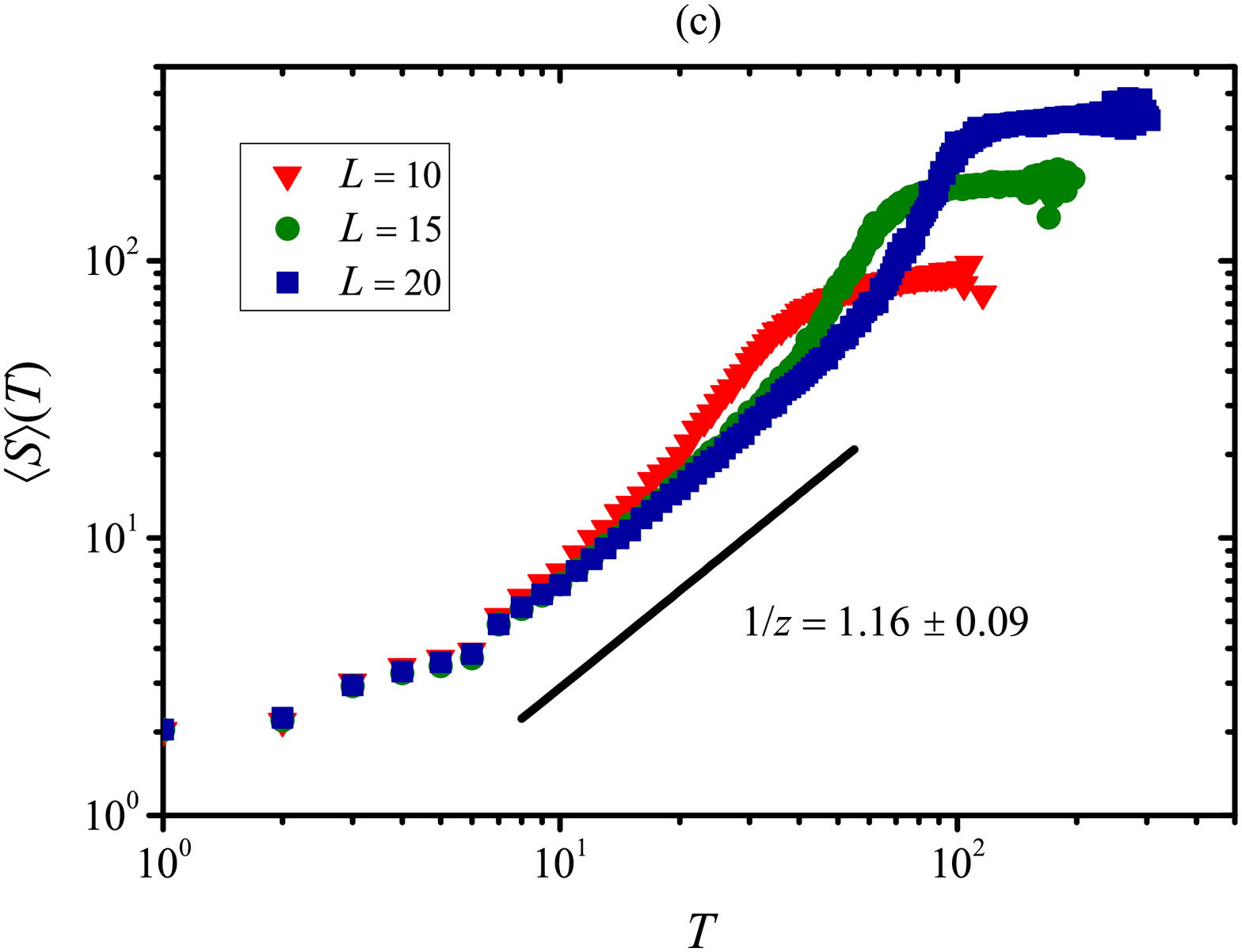}
\caption{ (Color online) Scaling of the cumulative probabilities of (a) avalanche size $P(S)$, (b) avalanche duration 
$P(T)$. (c) Scaling of the conditional average of avalanches with duration $T$. The results are obtained using $\gamma=0.001, \sigma=0.0001, S=0.001005$ at $K=0.0018$ and the lattices with the linear size $L=10, 15, 20$.}
\label{fig:f5}
\end{figure}

As predicted by renormalization group theory \cite{James}, avalanche data collapses onto universal scaling functions near a critical point and follow:

\begin{eqnarray} \label{fs}
p(S) &\sim& S^{-\tau},\nonumber\\ 
p(T) &\sim& T^{-\beta},\nonumber\\
\langle S \rangle (T) &\sim& T^{1/ { z}},
\end{eqnarray}
where $p(S)$ and $p(T)$ are the probability density functions of the avalanche size and duration, respectively. $\langle S \rangle (T)$ is the average of avalanche size conditioned on a given duration \cite{James}. $\tau$, $\beta$ and $1/ { z}$ are the critical exponents of the system and are independent of the details of the model or system \cite{Friedman2012}. The scaling theory requires the following relation between the exponents
\begin{equation} \label{s}
\ \frac{\beta-1}{\tau-1} = \frac {1} {z}.
\end{equation}
Mean field theory predicts $\tau=3/2, \beta=2.0$ and $1/ {z}= 2.0 $ based on SOC \cite{James}.

The results of our simulation for three lattice sizes $10\times 10$, $15\times 15$ and $20\times 20$ at $K_c=0.0018$ are illustrated in Fig.\ref{fig:f5}. In this figure, panels (a) and (b) show the scaling of the cumulative probabilities of the avalanche size ($P(S)=\int_{S}^{\infty} p(s) ds$) and avalanche duration ($P(T)=\int_{T}^{\infty} p(t) dt$), respectively, and panel (c) is devoted to the scaling of conditional average avalanche size, $\langle S \rangle (T) $ in terms of the duration $T$. According to equation \eqref{fs}, $P(S)$ and $P(T)$ scales as 
\begin{eqnarray} \label{fs2}
P(S) &\sim& S^{1-\tau},\nonumber\\ 
P(T) &\sim& T^{1-\beta}.
\end{eqnarray}

For the critical point suggested by temporal complexity, we found the following exponents: $\tau=1.61 \pm 0.13$, $\beta=1.70\pm 0.06$, and $1/ {z}=1.16 \pm 0.09$. At criticality, the scaling relation, Eq.\ref{s}, is met. This is another proof that the critical point suggested by temporal complexity is in fact the system's critical point. It is notable that we have considered $\sigma=0.0001$ throughout the text, and as illustrated in fig.\ref{fig:f4}, increasing noise, $\sigma$, comply increasing $K$, in order to keep these results consistence. 

\section{Concluding Remarks}
Here, by continuous solution of our previous model \cite{Zare2013}, and consequently using Gaussian noise with zero-mean, we explored if at the critical point indicated by temporal complexity, avalanche data collapses onto universal scaling function as predicted by the theory of dynamic critical phenomena \cite{James} and is consistent with the empirical founding of \cite{Beggs2003}. 
Our results provide compelling evidence that temporal complexity and neural avalanches both display power-law behavior at the critical point. Therefore, we may conclude that the choice of noise would result in the inconsistency with the result of the current work with the previous one \cite{Zare2013}. In fact, the continuous time description is more natural and make our model comparable to the empirical results.
We emphasize that the exponents of avalanche data collapses on scaling exponents that are model independent and identical for all systems in the same university class. On the other hand, the results of the paper indicate that a simple model using a regular lattice successfully sheds light into cooperation-induced criticality in neural system and establishes a connection between criticality and neural avalanches. For the future research, we will explore inhibitory connections in the network to make our model closer to real brain networks. 

In light of the results of the current research, we conclude that neural avalanches are indicators of criticality if continuous time description is applied to the LIFM model. This may also indicate that spatial scale invariance given by avalanche cumulative probability distribution in size displays relatively slow numerical convergence 
compared to temporal complexity measure. 

\section{Acknowledgement}

The authors gratefully acknowledge financial support from Iranian Cognitive Science and Technologies council through Grant No.2960.

\end{document}